\documentclass[12pt,aps]{revtex4}
\usepackage{epsfig}
\usepackage{amsmath,amssymb,color}
\usepackage{epstopdf}
\usepackage{graphicx}
\graphicspath{{pictures/}}
\DeclareGraphicsExtensions{.pdf,.png,.jpg}
\parskip=\medskipamount

\newcommand{\be}{\begin{equation}}
\newcommand{\ee}{\end{equation}}
\newcommand{\eq}[1]{(\ref{#1})}
\newcommand{\fig}[1]{Fig.\ref{#1}}

\newcommand\disp{\displaystyle}

\begin{document}

\title{Spectral peculiarity and criticality of the human connectome}

\author{K. Anokhin$^{1,2}$, V. Avetisov$^3$, A. Gorsky$^{4,5}$, S. Nechaev$^6$, N. Pospelov$^1$, and O. Valba$^{3,7}$}

\affiliation{
$^1$Lomonosov Moscow State University, 119991, Moscow, Russia \\
$^2$National Research Center "`Kurchatov Institute"', 123098, Moscow, Russia \\
$^3$N.N. Semenov Institute of Chemical Physics RAS, 119991 Moscow, Russia \\
$^4$Institute for Information Transmission Problems RAS, 127051 Moscow, Russia \\
$^5$Moscow Institute of Physics and Technology, Dolgoprudny, 141700 Russia \\
$^6$Interdisciplinary Scientific Center Poncelet (CNRS UMI 2615), 119002 Moscow, Russia \\
$^7$Department of Applied Mathematics, National Research University Higher School of Economics, 101000 Moscow, Russia
}

\begin{abstract}

We have performed the comparative spectral analysis of structural connectomes for various
organisms using open-access data. Our analysis indicates several new peculiar features
of the human connectome. We found that the spectral density of human connectome has the maximal deviation from the spectral density of the randomized network compared to all other organisms. For many animals \emph{except human} structural peculiarities of connectomes are well reproduced in the network evolution induced by the preference of 3-cycles formation. To get the reliable fit , we discovered the crucial role of the \emph{conservation of local clusterization} in human connectome evolution. We 
investigated for the first time 
the level spacing distribution in the spectrum of human connectome graph Laplacian. It turns out that the spectral statistics of 
human connectome
corresponds exactly to the critical regime familiar in the 
condensed matter physics which is hybrid  of Wigner-Dyson and Poisson distributions.
This observation provides the strong support for the much debated statement of the brain criticality.
\end{abstract}

\maketitle

\section{Introduction}
\label{s:01}

\subsection{Purpose of the work}
\label{s:01-1}

Understanding basic mechanisms of brain functioning in terms of the structure of underlying anatomical and functional brain networks, is a challenging interdisciplinary issue which worries researchers over the decades. Detailed presentation of a current state of the comprehensive studies of structural and functional neural connectivity referred as 
connectomics can be found in \cite{book,kaiser,sporns1, sporns2}.
To summarize the mainstream directions of modern research, one can highlight two questions of the primary interest:
\begin{itemize}
\item Which properties of the connectome are of key importance for an effective brain functioning at the cognitive level  and 
information processing?
\item What are the operational mechanisms of the structural network evolution, allowing to arrive at a present pattern of the connectome organization?
\end{itemize}

Answers to these questions are currently sought in the studies ranging from the investigation of the complete network of connections among the 302 neurons of the nematode Caenorhabditis elegans (C. elegans) \cite{chklov} to analysis of complex mammalian brain networks including the rat, cat, macaque and the human connectome \cite{book, cat,macaque,rat, odor}. The initial analysis of the C. elegans data has led to a conjecture that from the topological point of view, the connectome is an example of a “small world” network with a high clusterization and “short path” structure \cite{ws}. In some sense, such a topological network lies between the regular and completely random (Erdos-Renyi) topological graphs \cite{dorog, newman1, newman2}. It was suggested that high clustering coefficient determines efficiency of inter-module brain processes, while small average path length contributes to the embedding of large regions into a high-performance network and thus allows to connect system processes on different scales. It this respect small-world model fitted well to the properties of brain networks that combine high local connectivity with global information transmission. It linked processing of neural information on local and global levels with peculiar properties of the brain network architecture.

Such a line of reasoning was developed in \cite{scalefree} where the connectome is identified with the scale-free network (see also \cite{odor}).  Scale-free networks are characterized by a power-law degree distribution according to which the majority of nodes show few links, but a small number of hub-nodes have a large number of connections and ensure a high level of global network connectivity \cite{73}. Both small-world and scale-free architectures are considered to be attractive candidates for efficient flow and integration of information across the network \cite{74,75,76}.

However, it has been also recognized  that some properties of the connectome, like the hierarchical structure and the vertex degree distribution, cannot be explained in the frameworks of the "small world" paradigm. The hypothesis that brain networks exhibit scale-free topology became popular at the turn of the millenium, however, nowadays there are many evidences that connectomes on various anatomical scales deviate from networks with the scale-free vertex degree distribution. The emerging viewpoint is that the connectome realizes a new type of a network architecture.

To unravel these specific organizational  principles the comparative analysis of connectomes of different organisms \cite{80} is of extreme importance. It provides hints for the identification of key structural properties of the neuronal network, crucial for 
its integrative functions across a variety of specific neuroanatomical organizations. Since any global brain function is a collective effect it should be treated by appropriate methods capable to catch collective properties of underlying structural network. From this point of view, the comparative investigation of different connectomes undertaken in \cite{repka} via the analysis of their spectral properties, seems particularly promising (see also \cite{yama}). In our work we continue and develop this comparative 
spectral approach.

Through the ultimate wiring of the neural network is cellular, the analysis of the most of brain networks on this scale is currently complicated due to the lack of available experimental data. The human brain is made up of $8.6\times 10^{10}$ neurons \cite{78}, and current imaging techniques are far beyond resolving its microscopic connectivity.  
Only neuronal connectomes of comparatively primitive organisms, such as  P.pacificus worm and a C.elegans nematode, are currently available. The nematode connectome reconstruction had started in 1974 and lasted 12 years, despite it contains only about 300 neurons and several thousands of synaptic connections \cite{79}. Thus in a huge number of works the brain networks are studied at large and middle scales. The nodes of such networks are either "voxels" (cubic 3D areas containing hundreds of thousands of neurons), or whole brain regions.

For our work we used data from open sources. The data on the macaque connectome is limited to the only one network obtained in the Cocomac project \cite{macaque}. Data on C.elegans and macaque connectomes have been taken from the Open Connectome Project \cite{OCP} database. Data on human structural connectomes were extracted both from Open Connectome Project and Human connectome project databases \cite {UMCD}.

Classical methods of the theory of complex networks are now widely used in neurobiology \cite{book,kaiser,sporns1} . However, many commonly used metrics, such as betweenness centrality, efficiency, and others focus rather on the properties of individual nodes of the network than on the features of the network as a whole. In this paper our attention is focused on the global properties of the connectome. For these purposes, the methods of spectral graph theory are well suited. The main objects of interest in this case are the eigenvalues and eigenvectors of the matrices characterizing a graph, for example, adjacency matrix or Laplace matrix. The spectrum of such a matrix (i.e. a set of eigenvalues) is a unique identifier of the network, its peculiar "fingerprint". Knowing the whole set of eigenvalues and eigenvectors, it is possible to restore the original appearance of the network (with rare exceptions, which are described below). In the process of spectral decomposition, matrix elements corresponding to different elements of the network are mixed in a complex way, creating a "global" portrait of the network as a single object. For example, in the case of a graph represented as a Laplace matrix, its eigenvalues have a clear physical meaning: they represent the frequencies at which the graph would resonate if it was made of springs.

Using this data we demonstrate that various structural and spectral properties of connectomes of different organisms can be designed evolutionary under specific constraints starting from the undirected "null network state" which is a randomized version of an initial graph. 

The evolution of the network is carried out using the Metropolis algorithm. Without the loss of generality, it can be described as follows: taking the "null network state" as the input, the Metropolis algorithm attempts to make random changes to the network. If these changes occur in the "right direction", i.e. reduce the distance between the current and the desired state of the network in a pre-selected metric, they are accepted with probability one. If, to the contrary, the elementary rewiring moves the network away from the desired state (for example, reduces the number of triangular motives, though the purpose of evolution is to increase their number), it is accepted with some probability, exponentially decreasing with the size of "wrong" deviation. The chemical potential, $\mu$, plays the role of a parameter, governing the probability to which these "wrong" steps in the network evolution are allowed. The chemical potential approves its name by the function it performs: bringing analogy from physics, $\mu$ controls the amplitude of "thermal fluctuations" in the algorithm known as "simulated annealing": in the absence of thermal fluctuations, the network accepts only "positive changes" in its evolution along the landscape and might be easily trapped in a \emph{local} energy minima. The possibility of some "backward moves" allows the system to escape from local traps, thus helping the network to reach the true ground state.

One can easily understand the sense of an evolutionary algorithm by considering an Erdos-Renyi network as the system's 'null state'.
The constrained Erdos-Renyi network (CERN) of $N$ nodes evolves under the condition that the vertex degree in each node is conserved during the network's rewiring. The spectral properties of CERNs were thoroughly investigated in \cite{crit2}. The "driving force" of the network's evolution is the attempt to increase the number of closed 3-motifs (closed triad of links).

\begin{figure}[ht]
\centerline{\includegraphics[width=14cm]{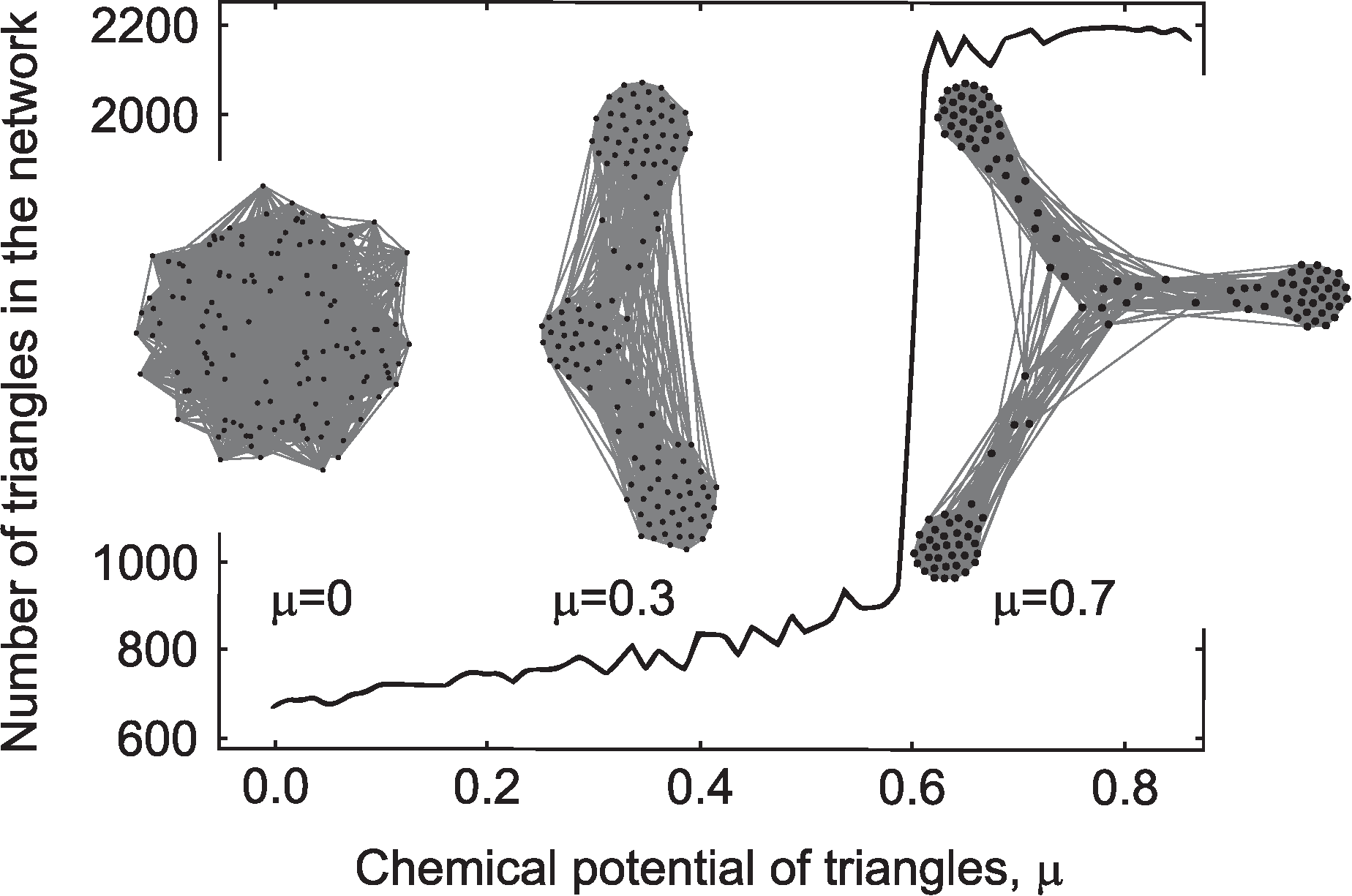}}
\caption{Phase diagram of the CERN. Number of triangles in the final network dramatically increases at the point $\mu = \mu_{crit}$ which for large networks corresponds to 1st order phase transition. Typical network structures are shown at three different values of $\mu$.}
\label{fig:01}
\end{figure}

The condition of the vertex degree conservation in each graph's node changes drastically the final state of the evolving structural network. These constraints provide $N$ hidden conservation laws for the stochastic network evolution (rewiring) making the corresponding dynamic system quite special. In particular, it was found in \cite{crit2} that the evolving CERN undergoes the phase transition and gets defragmented into a set of $K$ dense communities when the chemical potential, $\mu$, of closed 3-motifs exceeds some critical value, $\mu_{cr}$. The number of communities, $K$, depends on the density of the network at the preparation conditions and can be approximately estimated as [1/p], where $[...]$ designates the integer part of $1/p$ and $p$ is a probability to connect any two randomly chosen vertices of ER network. The phase diagram of the random constrained Erdos-Renyi network (with fixed vertex degree) and enriched by closed 3-motifs (controlled by the chemical potential $\mu$) is shown in \fig{fig:01}. To make the figure more informative, we show network samples at three different densities of closed 3-motifs. The effective way of constructing the phase diagram is discussed at length below.

The peculiarities of the network dynamics \emph{for humans} prompted us to suggest an existence of $N$ \emph{additional} conservation laws consisting in conservation of \emph{local clusterization} for each network node. The local clusterization has been used in the analysis of generic exponential graphs (see, for instance \cite{boguna}), however in the context of the connectome, it has been applied for the C.elegans only in rather restricted context. We shall use local clusterization as an "order parameter" which can distinguish the connectome of humans from connectomes of other animals.

In our work we tackle in details spectral properties of experimentally available adjacency and Laplacian matrices of structural networks. The simplest characteristics of the spectrum is its spectral density. However  more refined characteristics like the statistical correlators of the spectral densities carry a bunch of the additional information about the network properties. The investigation of the level spacing distribution (i.e. the distribution of distances between the neighboring eigenvalues) allows one to identify the level statistics using the standard methods of spectral statistical analysis. The level spacing distribution provides the key information concerning the localization properties of the signal propagation in the connectome. To the best of our knowledge, the spectral statistics has not yet been discussed in the context of the structural connectome.

Our statements are as follows:
\begin{itemize}
\item The spectral density of adjacency matrices for human structural connectomes has the maximal deviation from the spectral density of the randomized (via the Maslov-Sneppen procedure) network in comparison with the other  organisms in this study.

\item Spectral density of adjacency matrices of structural connectomes for all organisms \emph{except humans} can be satisfactorily reproduced by conserving the vertex degree in every network node and reconnecting links under the control of the chemical potential of triangles in the stochastically evolving network (i.e. restoring the initial connectome level of clustering in a randomized network)

\item Spectral density of structural  \emph{human} connectomes can be reproduced well enough by random rewiring of network's links if besides the conservation of the vertex degree, we demand also the conservation of the local connectivity (i.e. imposing extra conservation laws).

\item The level spacing distribution for  human connectomes demonstrates very peculiar behavior which
corresponds precisely to the critical regime and is the hybrid of Wigner-Dyson and Poisson level 
statistics which means that the human connectome is at criticality. This differs from the model of the cluster-enriched scale-free network.

\end{itemize}

\subsection{Structural and dynamic properties of a connectome}
\label{s:01-2}

Let us summarize key statistical and dynamic properties of connectomes of various organisms on the basis of open-sources data analysis. These properties we consider as the reference point for our investigation. Firstly, connectomes typically have a large modularity. Recall that the modularity measures the clusterization of the network and reflects its hierarchical structure \cite{spornshub, harriger, bullmore1}. The large modularity is confirmed in \cite{50000} by the high-resolution analysis of adjacency matrices of human connectome for the $N>5\times 10^4$ nodes. Secondly, structural connectomes have the "short path" property typical for the small-world networks. Thirdly, spreading of the signal in connectomes demonstrates the synchronization \cite{gollo1,gollo2,brainsin}.

Important property of the connectome is the distribution of so-called "motifs" -- small subgraphs of the whole network. For C.elegans this quantity has been studied for the first time in \cite{spornsmot}, where it was found that the number of open 3-motifs (connected pairs of network links) significantly exceeds the respective quantity for the random Erdos-Renyi network with the same link formation probability. The situation with closed 3-motifs (fully connected triads of links) is more subtle: First, it was suggested that there is no exceed of the number of closed 3-motifs \cite{spornsmot}. However later, a more detailed analysis showed that the connectome still is enriched by the number of closed triads of links compared to an Erdos-Renyi network of same density. Moreover, the connectome has a preferred typical topology of a backbone between communicating neurons \cite{pathmotif}.

It has been shown in \cite{spornsmot,song} that all connectomes have an excess of open 3-motifs and the directed open 3-motifs are highly important. In particular  it was found that substituting open 3-motifs by closed 3-motifs, one can sharply decrease the synchronization of the connectome which means that presence of open 3-motifs is crucial for the synchronization. It was argued in \cite{anton} that the evolution of the connectome from the C.elegans to the human shows the improvement of the flow propagation in the network.

In \cite{pathmotif} the authors have investigated the distribution of "paths motifs" in  connectomes. The following topological classification of paths has been adopted: L-paths between the nodes with a mean degree, R-paths between the nodes with a mean degree and a hub, G-paths between the hubs. It was found that the most often path in a human connectome has a L-R-G-R-L motif.

Much information concerning the structural properties of the connectome is stored in the spectral properties of corresponding adjacency and Laplacian matrices. The spectral density of the Laplacian matrix of  the connectome of C.elegans is the triangle-shaped "continuum" zone accompanied by several low-energy isolated eigenvalues \cite{repka}. Such a shape of the spectral density is far from the one for random Erdos-Renyi random graph, which has the oval-shaped Laplacian spectrum. The low-lying eigenvalues and corresponding eigenfunctions of the Laplacian matrix of the connectome carry the information about the transport properties in the network. In particular, recently it has been found  (see \cite{hemi}) that the second eigenvalue, $\lambda_2$, is responsible for the diffusion of the signal between two hemispheres. The third eigenvalue, $\lambda_3$, seems to measure the radial diffusion in the connectome from the inner to outer regions \cite{hemi}. The largest eigenvalue of the connectome Laplacian does not deviate much from the eigenvalue typical for a purely random network with the same averaged characteristics, which means that the connectome typically does not develop the bipartite structure.

\section{Modelling of connectome evolution by motif-driven rewiring in constrained 'null-state' network}
\label{s:02}

\subsection{Spectral density of networks and motif-driven network evolution}
\label{s:02-1}

Here we describe the numerical procedure which manipulates by the experimental data on structural connectomes taken from open sources. Our experiment is aimed to reveal the principle "conservation laws" which might govern the structural transformation of the connectome during the biological evolution.

Before proceeding further, one important remark is appropriate. There is a common belief supported by many numerical simulations , that the eigenvalue density (spectral density) of a graph adjacency matrix is a "fingerprint" of a corresponding network in generic situation. Besides, there are known examples of "iso-spectral" graphs which have different adjacency matrices, however their spectra coincide. Such situations are rather exceptional and practically do not occur in randomly generated patterns (their Kolmogorov complexity is very high). In our study we consider the spectral density as a graph invariant which sets a "metric" for graphs: if spectral densities of two graphs are similar, we say that the adjacency matrices are similar, while as less two spectral densities resemble each other, as more unlike the graphs are. For the quantitative comparison of spectral densities we use (i) the "transport metric" (see \cite{emd2} for precise definition). Thus, rewiring the network, we catch the evolution of the corresponding spectral density.

All the network spectra plots below were constructed by simple convolving of the set of eigenvalues with a Gaussian curve and further normalization to make the area under the distribution equal to 1.

The setting of the simulation is as follows. We take structural connectomes (the state $S$) of C.elegans, macaque and human, defined by adjacency matrices of corresponding networks, and destroy the network patterns by random rewiring of links under the condition of the vertex degree conservation at each graph node, thus getting the state $S_{rand}$ (the "null state"). To obtain null-state networks, we used the Maslov-Sneppen randomization (MS) algorithm \cite{maslov} (see Fig.2). The rewiring procedure retains the size of the network and its density, and also strictly preserves the degree of all nodes. Degree distribution was shown to be of key importance for the network's structure, that is why we consider MS-randomized networks as "null-state" patterns. Despite the vertex degree distribution of randomized ($S_{rand}$)- and initial ($S$)-networks is the same, their topological, motif, spectral and other properties can be essentially different.

Now, starting from the state $S_{rand}$, we are trying to recover back the $S$-state by continuing random rewiring of links (again with the vertex degree conservation), however now -- under the influence of a "driving force" via the Metropolis algorithm described above. The closeness of two network states, the initial and the randomized one, is measured by the distance between their spectra.

\begin{figure}[ht]
\centerline{\includegraphics[width=16cm]{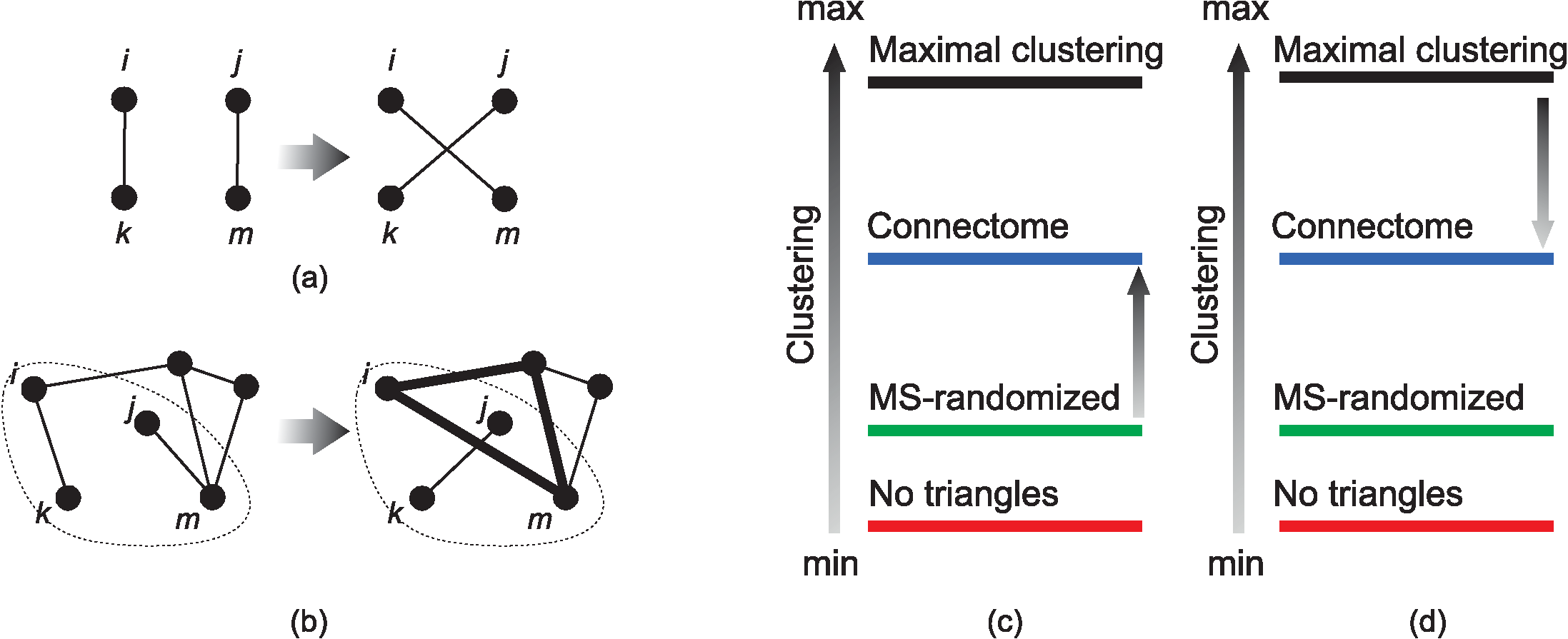}}
\caption{(a) Single rewiring step of Maslov-Sneppen algorithm preserving degrees of all vertices; (b) Example of a local network updating move which increases the number of triangles; (c) Maslov-Sneppen randomized network, exposed to evolutionary process, which slowly increases the network clustering until it reaches the initial level of the connectome network; (d) Maximally clustered network, exposed to Maslov-Spappen randomization algorithm, resulting in a network with the same clustering. Arrows indicate the directions of the evolutionary process.}
\label{fig:02}
\end{figure}

The standard Maslov-Sneppen Metropolis algorithm described above is transparent and straightforward in implementation. However, performing evolution of MS-randomized network to a highly clustered state, requires a lot of computational resources: it takes much more time to pull the network evolution towards a state with a given density of closed 3-motif state, than to push it to the "null state". To reduce significantly the time of computations, we have implemented another procedure which gives the same result. Instead of starting from a completely randomized network state with the preserved degree distribution, we have constructed a \emph{maximally clustered} network (MCN) (again respecting the vertex degree conservation). The MCN is a graph that has maximal (or nearly maximal) number of closed 3-motives available for a given degree distribution.
Thus, the initial network is fully reorganized in the process of MCN construction. Taking MCN as a "null state" and running the Maslov-Sneppen Metropolis procedure controlled by the density of closed 3-motifs, we bring the network to desired clustering level. Such a technical trick allows to perform computations for highly clustered graphs very efficiently. Figuratively, one can say that assembling the network with specified density of closed 3-motifs from the randomized state is like a "rising to the mountain", while assembling the network from maximally clustered state is like a "descent from the mountain", which is less energy consuming. Schematically we have depicted these two algorithms in the \fig{fig:02}.

So, we have two types of "null-state" networks: the MS-randomized network and MCN. Selection between them depends on the choice between the clustering evolution in "up" or "down" directions. In our work we use two types of additional constraints imposed on the null-state network evolution: 

\begin{itemize}
\item The "first level", the triangle-preserving constraint (TPC), is the condition of maximizing the number of closed 3-motifs in the evolving graph until its clustering reaches the level of the initial (pattern) network. Using this procedure, we calculate only overall number of triangles, while we do not pay attention to how these triangles are distributed over the network. In general, such a "total clustering preserving procedure" allows to get a network state which is closer to the original one than the simple MS-randomized network.
\item The "second level", the local clustering constraint (LCC), preserves local clustering \emph{in each node} of the network such, that the rewired network (besides to vertex degree conservation), also preserves the number of closed 3-motifs in each node. The LCC algorithm is thoroughly described below.
\end{itemize}

\subsection{Network randomization with additional conservation laws}
\label{s:02-2}

More refined algorithms of network rewiring are demanded to preserve simultaneously several characteristics of the original graph. We are guided by an attempt to propose the "minimal" model which, on one hand, could capture the key properties of the structural connectome and could distinguish humans from other organisms, and on the other hand, is as simple as possible. We put forward the conjecture that such additional "feature" that should be conserved during the network randomization together with the vertex degree conservation, is the number of triangles in which each node participates.

Rephrasing the said above, we randomize the network, however conserve in all nodes: i) the vertex degree; ii) the clustering coefficient. Below we propose our own way to solve this problem numerically using a modification of the Metropolis algorithm. Having the network adjacency matrix, $A=\{a_{ij}\}$, of a graph ($(i,j) = 1,...,N$), consider two auxiliary diagonal matrices, $D_{\star}$, $D_{\triangle}$, defined as follows
\be
\begin{cases}
\disp D_{\star}= \{d_1,d_2,...,d_N\}: \quad d_i = \sum_{j=1}^N a_{ij} \\
\disp D_{\triangle}=\{\tilde{d}_1,\tilde{d}_2,...,\tilde{d}_N\}: \quad \tilde{d}_i = \sum_{j \neq k}^N a_{ij} a_{jk} a_{ki}
\end{cases}
\label{eq:dd}
\ee
The elements of the matrix $D_{\star}$ are the degrees of nodes, while the elements of the matrix $D_{\triangle}$ are the numbers of triangles into which a specific node is involved. From the general point of view, our algorithm resembles the simulation of some physical process that occurs when a substance crystallizes. It is assumed that the "crystal lattice" has already been formed, however the transitions of "individual atoms" from site to site are still permissible.

It is assumed that the destination state of the network is the configuration in which clustering of all nodes is the same as in the "pattern state" (there can be many such destination networks). Our algorithm takes as an input the network in which the information about the clusterization is "washed out". Then we rewire the network, conserving all the degrees of nodes. After each rewiring, we compute the "distance" between the resulting modified network and the preselected pattern network
\be
F=\sum_{i=1}^N\left|C_i-C_{i0}\right|
\label{eq:F}
\ee
where $N$ is the number of nodes of the network, $C_i$ is the clustering coefficient of the node $i$ in the evolving network, while $C_{i0}$ is the clustering coefficient in the pattern network which the evolving network tends to reach. Another definition of the "distance" which sets the metric in the space of the "networks similarity" is as follows
\be
T=\sum_{i=1}^N|T_i-T_{i0}|
\label{eq:T}
\ee
This definition is equivalent to \eq{eq:F}, except that the clustering coefficient is replaced by $T_i$, the absolute number of triangles involving the vertex $i$.

The Metropolis-like algorithm trying to minimize the cost functions $F$ \eq{eq:F} and $T$ \eq{eq:T} is set as follows. First, we chose the metrics \eq{eq:F} or \eq{eq:T}. If the local random perturbation (rewiring) of the network is such that the system tends to the desired destination state (i.e. $F$ or $T$ decrease), this random step is accepted with the probability 1. Otherwise, if $F$ or $T$ are increased by $\Delta\ge 0$ in the selected metric, the corresponding step is accepted with the probability $e^{-\mu \Delta}$, where $\mu>0$ is the chemical potential of Metripolis procedure. After reaching the local energy minimum (which is calculated on the basis of the best time/distance ratio), the algorithm updates the metric and repeats the procedure. We have mentioned that such an algorithm known as the "simulated annealing", allows to reach the ground state of the system without getting trapped at local minima in the very complex energy landscape.

\begin{figure}[ht]
\centerline{\includegraphics[width=14cm]{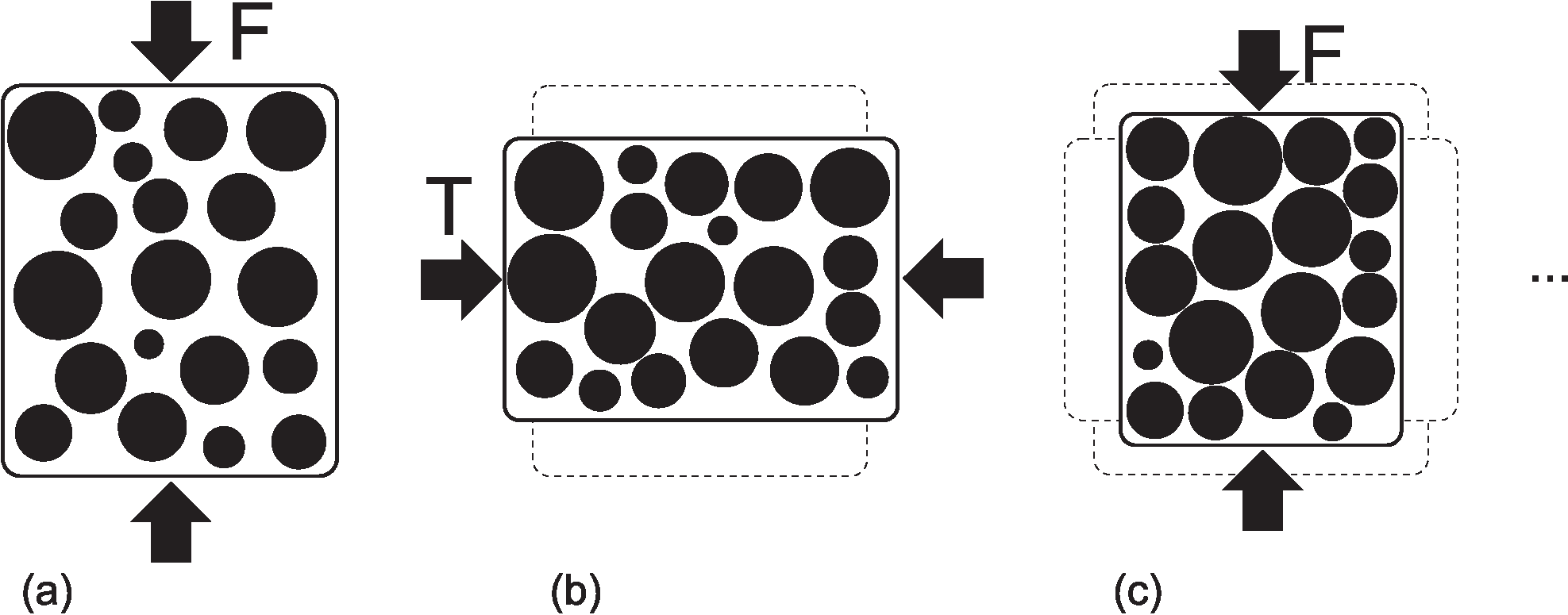}}
\caption{Illustration of a repetitive sequential minimization of $F$ and $T$. Minimizing $F$ we compress the pile, however can slightly increase $T$. Minimizing $T$ afterwards, we compress the pile further, however could slightly increase $F$. Repeating $F-T-F-...$ compression and shacking randomly the pile, we reach the densest packing.}
\label{fig:03}
\end{figure}

One can say that computing the "distance" from the evolving network to its final destination, via the clustering coefficients, $F$, we "equalize" nodes with different vertex degrees. Namely, for all nodes the clustering coefficient lies between 0 and 1. Thus, nodes with low degrees, "pull" the triangles from hubs during the network evolution as they are widely spread. On the other hand, the rewiring procedure via $T$-metric satisfies the interests of large nodes to the detriment of small ones. We have developed the optimization procedure in which both metrics $F$ and $T$ are simultaneously used. The algorithm sequentially switches between these metrics and adjusts the network for both $F$, respecting the interests of loosely connected nodes, and $T$, which works well for hubs. Schematically the idea of the algorithm is depicted in \fig{fig:03} on the example of sphere packing. Vertical compression (associated with the minimization of $F$) of the random 2D pile of spheres leads to a desired increase of the density, however might be accompanied by the increase of a horizontal size (associated with $T$). To squeeze the pile more, we shake the pile randomly and compress it in the horizontal direction, then we switch back to the vertical compression, etc, until the densest packing state is reached.

Arriving at the stationary state, when $F$ and $T$ cannot be decreased anymore during reasonable time, we compare the adjacency matrices of the destination network with the preselected pattern. This can be done by comparing the corresponding spectra of two matrices. The distance between spectra (which, by virtue of the above comments about the uniqueness of the spectrum, is understood as a quantitatively expressed degree of the "dissimilarity" of the two networks) is measured in terms of the so-called "earth mover's distance" metric (or Wasserstein's metric). The earth mover's distance (EMD) is a metric based on the minimum cost of transforming one histogram into another. Representing two distributions (two spectral densities) as two heaps of earth that need to be superposed by transferring small pieces of earth, then EMD determines the least amount of work required to accomplish this task. The calculation of EMD is based on solving the transport linear programming problem, for which effective algorithms are known. In our work we were using an open-source Python package designed for fast EMD computation \cite{emd1, emd2}.

\section{Results}
\label{s:03}

\subsection{Significant non-randomness of a human connectome compared to other organisms}
\label{s:03-1}

\begin{figure}[ht]
\centerline{\includegraphics[width=16cm]{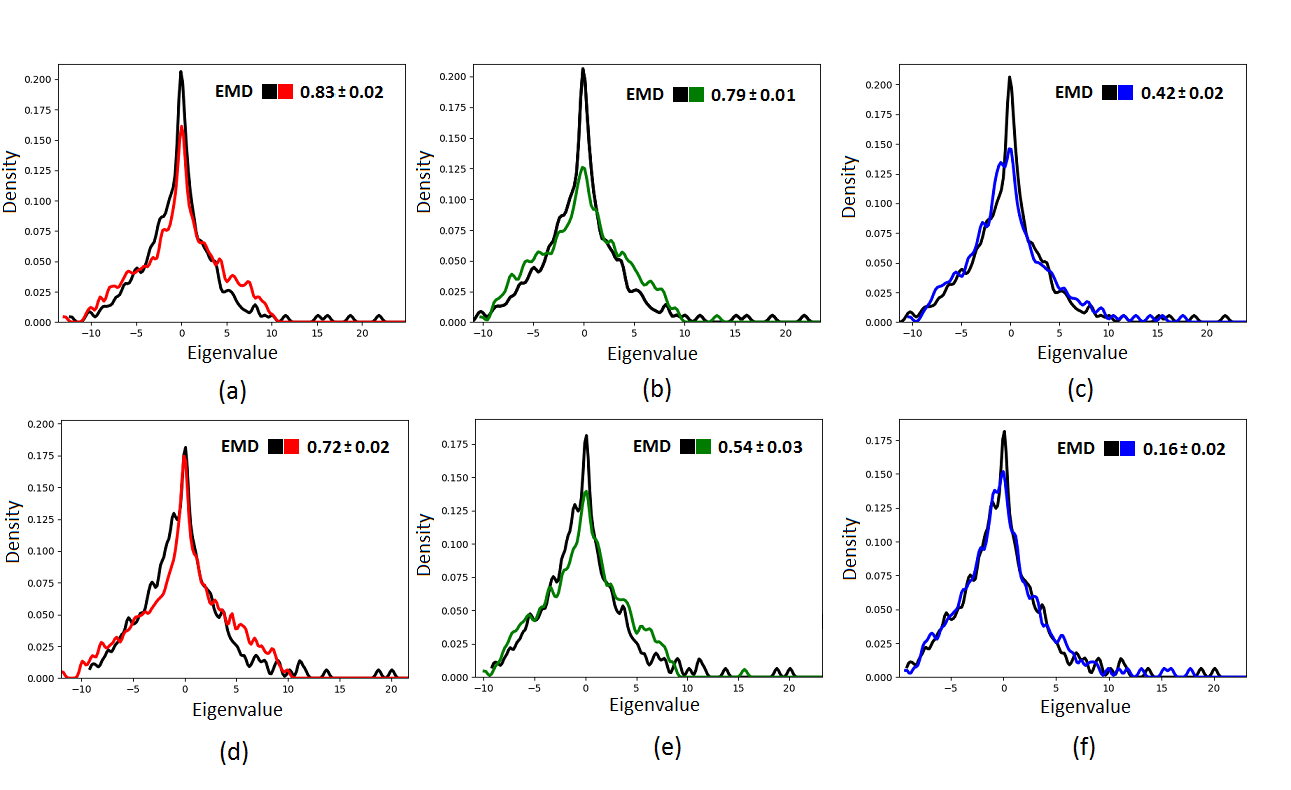}}
\caption{Spectra of adjacency matrices of a animal connectomes along with its MS-randomized version (red), TPC-randomized version (green) and LCC-randomized version (blue) (a)-(c): Macaque connecome, (d)-(f): Nematode connectome; MS = Maslov-sneppen, TPC = triangle preserving constraint (first evolutionary algorithm), LCC = local clustering constraint (second evolutionary algorithm)}
\label{fig:04}
\end{figure}

We have compared spectral densities of adjacency matrices of structural connectomes of various organisms with their Maslov-Sneppen randomized "null states". Our numeric analysis presented in \fig{fig:04} allows to conclude that the "null state" network with the vertex degree conservation constraint, reproduces with a good accuracy the spectrum of initial connectome adjacency matrix of C.elegans, macaque, \emph{but not of a human}.

It means that the vertex degree conservation is a candidate for a "sufficient" minimal set of conservation laws which control the evolution of these organisms. Imposing the "first level" additional constraint (i.e. preserving overall number of triangles in the network) we may slightly improve the spectra coincidence for nematode and macaque and significantly---for a human. However, the spectral distance between the human connectome and its TPC-randomized version is still essential, indicating that preserving overall clustering is not sufficient to reproduce the network properties.

To have the quantitative characteristics of the difference between spectral densities, we have
computed the earth mover's distance (EMD), $E$, between spectral density of original network and its MS-randomized analog, conserving vertex degrees in all network nodes. We found that EMD for a human connectome, $E_H$, is much larger than for connectomes of other animals:

\begin{table}
\begin{center}
\begin{tabular}{|c|c|}
\hline
{\bf organism} & {\bf EMD from randomized "null state"} \\ \hline \hline
human (1) & $E_{H1} = 1.45 \pm 0.02$ \\ \hline
human (2) & $E_{H2} = 1.53 \pm 0.02$ \\ \hline
macaque & $E_M = 0.83 \pm 0.02$ \\ \hline
nematode & $E_N = 0.72 \pm 0.02$ \\ \hline
\end{tabular}
\end{center}
\caption{Comparative analysis of animals by their earth mover's distance (EMD) from MS-randomized "null state" network.}
\label{t:SD}
\end{table}

The results presented in the table \label{t:SD} signal that the human connectome is much farther from the MS-randomized "null state" than the neuronal networks of other animals. One could speculate that such a difference is the consequence of the evolutionary selection acting on the neuronal network. This issue will be discussed in more details in the Discussion.

To test whether the stated results are not the artifacts of the network size (the investigated human connectomes had about 1000 and 600 nodes respectively, while networks of other animals had less than 300 vertices), we have performed numerical experiments on smaller neuronal networks. For the experimental data taken from the "Open Connectome Project" we obtained networks of different size (from 250 to 3000 nodes) and found that the difference in the spectral distance does not change during such scaling. We have also carried out numerical experiments with the data on human connectomes taken from other researches (data available at UMCD database, graph construction algorithms do not coincide) to exclude the possibility that our result is an artifact of algorithms used in the data processing of databases. The effect of increasing the spectral distance for humans with respect to other animals is supported, what is an indirect proof of its generality.

\subsection{Impact of local clustering on the network spectrum}
\label{s:03-2}

We have seen in the previous Section that the evolution of a human connectome is much more complicated compared  to other considered organisms, and to restore back the spectrum of the human structure network pattern from its MS-randomized "null state" requires some extra constraints (conservation laws). The "second level" algorithm preserving local clustering permits to advance in reproducing the connectome structure of humans. Some properties of exponential graphs with such set of local constraints were discussed in \cite{boguna}.

Among the characteristics of the network that affect its spectrum, the number of triangles, $T_i$ involving some given node, $i$ ($i=1,...,N$) is of much importance. Rephrasing that, one can say that the impact of the local clustering coefficient associated with a given node $i$, is crucial. Conserving these $N$ additional quantities $\{T_1,...,T_N\}$  (for all network nodes), one can significantly improve the coincidence of the spectra of the pattern and MS-randomized networks of human structural connectome in terms of the earth mover's distance, $E_T$, as it is shown in \fig{fig:04}.

\begin{figure}[ht]
\centerline{\includegraphics[width=16cm]{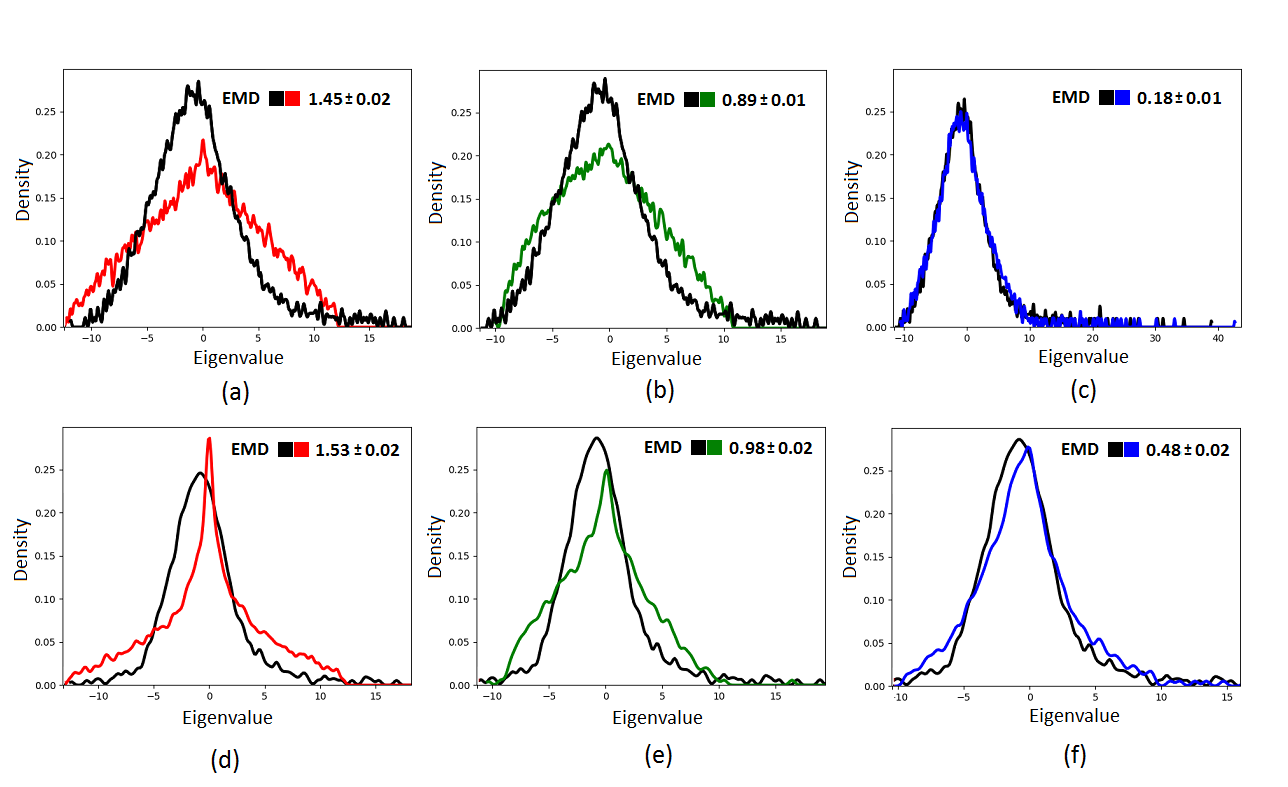}}
\caption{Spectra of adjacency matrices of a human connectome along with its MS-randomized version (red), TPC-randomized version (green) and LCC-randomized version (blue) A. Data from Hagmann(2008) B. Data from Open Connectome project; MS = Maslov-Sneppen, TPC = triangle preserving constraint ("first level" evolutionary algorithm), LCC = local clustering constraint ("second level" evolutionary algorithm).}
\label{fig:05}
\end{figure}

One of the main conclusions of our work is as follows. The coincidence of spectral densities between structural network of a human and its MS-randomized version exposed to evolutionary process could not be achieved by preserving only the \emph{average} clustering of the total number of triangles. Instead, the full vector $\mathbf{T}=\{T_1,...,T_N\}$ for all network nodes should be conserved. This result is stable for all organisms under investigation and for networks of various sizes. It is assumed that such a feature can be associated with the important role of local clustering in the structure of neuronal networks.

Let us recall that the similar set of conservation laws has been used in analysis of real networks in \cite{boguna}. The connectome of the C.elegans has been used as one example and it was argued that tuning the single parameter which controls the local connectivity, is possible to fit well the spectral density. Our study provides further evidence of the importance of various conservation laws in the evolution of connectomes.

\subsection{Criticality of the human connectome}
\label{s:03-3}

Let us begin with some definitions. We have defined already $A = \{a_{ij}\}$ -- the adjacency matrix of an undirected network (i.e. $a_{ij}=a_{ji}$). The  matrix elements, $a_{ij}$, take binary values: $a_{ij}=1$, if the monomers $i$ and $j\neq i$ are connected, and $a_{ij}=0$ otherwise. The absence of self-connections means that the diagonal elements vanish, i.e. $a_{ii}=0$. At length of the current work, we have studied spectral properties of adjacency matrices of networks, however in many papers another characteristic, the Laplacian of the graph, is under the investigation. The Laplacian matrix, ${\cal L}$, of a network is, by definition, 
\be
{\cal L}=d I-A
\label{eq:L}
\ee 
where $d$ is the vector of vertex degrees of the network, and $I$ is the identity matrix. The eigenvalues, $\lambda_n$ ($n=1,...,N$) of the Laplacian ${\cal L}$ are all real. For regular graphs (i.e for graphs with constant vertex degrees) the spectra of $A$ and ${\cal L}$ are connected by a linear transformation. 

The spectrum the of Laplacian ${\cal L}$ is positive and the minimal eigenvalue, $\lambda_1$, is zero. From the graph theory it is known that the multiplicity of the lowest eigenvalue, $\lambda_1=0$, equals to the number of disconnected components in the network. This fits with the identification of the number of separated discrete modes as of the number of clusters. Indeed, when some isolated eigenvalue hits zero, the cluster becomes disconnected from the rest of the network. The second eigenvalue, $\lambda_2$, carries the essential topological information about the network, known as the "algebraic connectivity", which measures the minimal number of links to be cut to get the disconnected network. The value of $\lambda_2$ plays an important role in relaxation and transport properties of the network, since it defines the inverse diffusion time, and plays crucial role in determining synchronization of multiplex (multilayer) networks \cite{arenas1}. The corresponding eigenvector (the so-called "Fiedler vector") sets the bijection between the network layers.

It is known that one of the most informative characteristics, which  delivers the information about the localization 
properties of excitations on the network, is the so-called "level spacing" distribution function, $P(s)$, where $s$ is the 
normalized distance between nearest-neighboring eigenvalues of the Laplacian matrix of the network. It is known from the classical theory of random matrices (see, for example, \cite{mehta}) that if $P(s)$ shares the Wigner-Dyson level statistics, the excitations are delocalized, while if $P(s)$ is exponential, the Poisson-distributed excitations are localized and the system behaves as an insulator:
\be
P(s)\sim \left\{
\begin{array}{ll} s e^{-s^2/\sigma^2} & \mbox{Wigner surmise (delocalized behavior)} \medskip \\ e^{-s/\delta} & \mbox{Poissonian statistics of events (localized behavior)}
\end{array} \right.
\label{eq:level}
\ee
where $\sigma$, $\delta$ are some positive constants, and $s=\frac{\lambda_i -\lambda_{i+1}} {\Delta}$ is the normalized gaps between nearest-neighboring eigenvalues.

However, there is the third critical regime for $P(s)$ which occurs when some control parameter is tuned exactly 
at the critical point and the system is at the point of phase transition. In this case the function $P(s)$
is hybrid of Wigner-Dyson and Poisson statistics at all energies. It has small-s behavior of the former and the large-s behavior 
of the latter one \cite{shklov}.  This hybrid statistics serves as the spectral mark of the criticality in the system.

\begin{figure}[ht]
\centerline{\includegraphics[width=14cm]{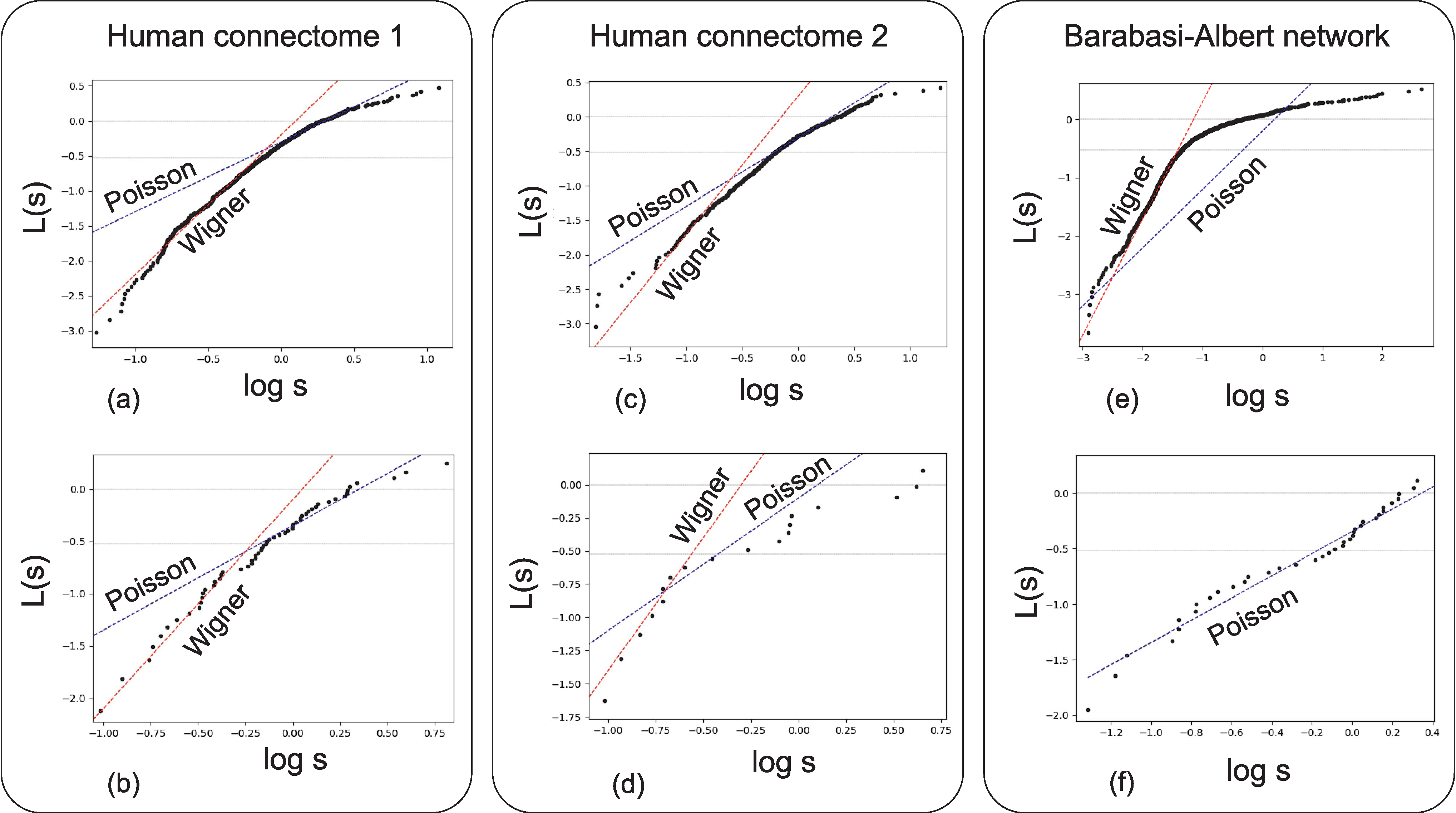}}
\caption{Level spacings of the Laplacian spectra in $(\log s,L(s))$-coordinates: (a-b) data set of human connectome human-1; (c-d) data set of human connectome human-2; (e-f) Barabasi-Albert network. Slope=1 (blue dashed curve) indicates the Poissonian statistics, slope=2 (red dashed curve) indicates the Wigner-Dyson regime.}
\label{fig:06}
\end{figure}

We have used a standard procedure to construct the level spacing: selecting a certain spectral region, $\Delta$, we computed a set of gaps between sequential eigenvalues and averaged them over $\Delta$. Finally, we presented the distribution of gaps between adjacent eigenvalues (in relative units) in coordinates $(x,y)$, which allows for straightforward identification of the spectral statistics by the slope of the curve: for the $x$-axis we have $\log s$, while for $y$-axis we have the function $L(s)$, defined as follows
\be
L(s)=\log \left(- \log(1 - C(s))\right); \qquad C(s)=\int_{-\infty}^s P(s')ds'
\label{eq:L}
\ee
where $C(s)$ is the cumulative distribution function of $s$. The main question is whether the level spacing distribution of human connectomes obeys the Wigner surmise, i.e. demonstrates the level repulsion, typical for interacting chaotic systems, shares the Poisson statistics, which means that the eigenvalues are uncorrelated or enjoys criticality? The results of our computations for Laplacian matrices of human connectomes  are presented in \fig{fig:06}.

The algorithm of computations is as follows. As one can see from \fig{fig:05}, the spectral density of adjacency matrix, $A$, of human connectome consists of a continuous (central) zone and a set of separated peaks (discrete zone, one separated eigenvalue per one cluster). In the Laplacian $L$, defined in \eq{eq:L}, one can also split the spectral density into continuous and discrete parts. In each such part we can determine the intervals $\Delta$, within which the level spacing belongs either to delocalized (Wigner), or to localized (Poissonian) subparts. It turns out that two regimes in  $P(s)$ both in continuous and discrete part of the spectrum
are separated by the crossover which corresponds to transition from Wigner to Poissonian statistics. The level spacing of two human connectomes are shown in \fig{fig:06}(a,c) for the continuous part of the spectrum, and in \fig{fig:06}(b,d) for the discrete one. In \fig{fig:06}(e) we have plotted for comparison the level spacing of Barabasi-Albert network Laplacian in the discrete part of the spectrum. No crossover is seen and all eigenvalues are localized.

Thus the spectrum of Laplacian matrices of human connectomes demonstrates 
a bit surprisingly  "hybrid" behavior for the level spacing distribution in all parts of the spectrum 
with a clear-cut crossover from Wigner-Dyson to Poisson behavior in each part of the spectrum as a function of the energy resolution $s$. 
This is exactly the universal critical behavior of $P(s)$  discovered at the edge of the Anderson localization  in \cite{shklov} which means that human connectome is at criticality.
The conjecture of the human connectome criticality is  controversial and highly debated issue (see \cite{criticality1,criticality2} for the recent discussions). Certainly, this conjecture is very attractive since in this regime we naturally have long-wave excitations
which exist for certain. Our result provides a strong support for the criticality conjecture from the standard  spectral
analysis viewpoint. Note that from our result is also clear that the 3d nature of the brain is essential since 
there are no localization/delocalization critical behavior in one and two space dimensions.

\section{Discussion}
\label{s:04}

\subsection{Main conclusions}

The idea of randomizing a network with preserving degrees of nodes (the Maslov-Sneppen algorithm) is not new, however in the literature it has been used typically outside the context of the spectral graph theory, being applied mainly to the determination of the average path length in the network, the global clustering coefficient, etc. In our work, following the ideas developed in \cite{crit2}, we use the Maslov-Sneppen randomization in combination with the spectral analysis. This allows us to uncover some hidden structural properties of network samples and analyze the stability of the spectrum with respect to the network perturbation. We have proposed the procedure to identify differences in the architecture of the connectomes of the organisms which stay on different steps of the evolutionary staircase.

The results of our study clearly demonstrate that some fundamental properties of the connectome cannot be explained by the behavior of typical network characteristics, such as the vertex degree distribution, the averaging clustering coefficient and the distribution of clustering among network nodes. This is especially true for human connectomes for which we have shown that the "earth mover's distance" (EMD) between the structural network pattern (represented by the adjacency matrix) and the corresponding randomized "null state" network, is essentially larger than respective EMDs for other animals. Such an interpretation raises a natural question about the significance of differences in the spectra of macaque and human connectomes, which evolutionary are much closer to each other than, the nematode. Yet the answer to this question is presently complicated due to the heterogeneity and small sets of available data.

We have performed crosscheck of our results on various sets of data to avoid the artifacts of specific algorithms of neurobiological data available from open sources. For this purpose we have used different human connectome data sets obtained by various experimental methods. Also, the sizes of considered networks varied from several hundred to several thousand nodes. However, for the macaque connectome such precautions are not yet possible, since the CoCoMac project data is the only complete source of information for the brain connectivity of this species.

The distribution of triangles for each vertex of the network seems to be a simplest invariant preserving the shape of the network spectrum. Keeping only an average clustering coefficient, we are unable to restore the spectrum of the network from its randomized version. For better reconstruction of the network topology it is not sufficient to know how many triangular motifs it has, but it is crucial how these triangles are distributed among the nodes. The vectors of triangular motifs, $\mathbf{T}$, in the real network pattern and in its randomized versions are not identical. Apparently, knowing $\mathbf{T}$ is crucial for reconstruction a modular brain architecture with several coupled hierarchical levels.

The importance of local clustering has been repeatedly emphasized in the  analysis of brain networks \cite{80,81}. It is suggested that  the combination of high local connectivity and the "small world" property on a large scale is responsible for many features of a brain functioning \cite{82}.  Perhaps, the local clustering should be considered as a crucial achievement of evolutionary selection, which essentially distinguishes the connectome from its randomized version.

We have provided  analysis of the eigenvalue correlations in spectra of Laplacian matrices of structural connectomes. Surprisingly
the level statistics turns out to be critical. This finding strongly support the widely discussed highly 
controversial brain criticality conjecture formulated long time ago \cite{formulation}. There are many arguments
in favor and against  this conjecture and our  spectral analysis yields one additional rock at the 'yes' side of the balance.

In our study we used the undirected structural networks which certainly restricts the reliability of our findings. Nevertheless, even for non-oriented structural connectome the spectral analysis provides the new important insights. Recent studies \cite{oriented} of oriented structural connectome show that new interesting features emerge which definitely deserve elaborations of new reliable mathematical models and clear physical explanations.

The issue of uniqueness of the human brain in it’s cognitive abilities and conscious information processing has been addressed at various levels of analysis, including evolutionary expansion of selective regions of the cerebral cortex, emergence of specific properties in the human neocortical neurons, novel kinds of cellular interactions, new molecular pathways, specificity of gene expression in neuronal and glial cortical cells \cite{66,67,68,69,70,71}. Our data contribute yet another dimension to this analysis by pointing at the peculiarity of the human connectome global organization. Its spectral characteristics support an expanded range of criticality known to maximize information transmission, sensitivity to external stimuli and coordinated global behavior typical of conscious states \cite{83,84,85,86,87}.

\subsection{Directions for further research}

Besides the consideration of oriented networks, the challenging question deals with understanding of interaction and synchronization of various functional sub-networks in the connectome. The first step on this way consists in the consideration of a two-layer network where there is a competition between strength of open in-layer 3-motifs and cross-layer pairwise interactions. In \cite{maslovpt} it has been found that within such a model one can see the phase transition between the phases with dominance of in-layer-- or cross-layer--connections. Depending on the parameters of the model, these two phases can be separated either by the sharp boundary or can be transformed one into another smoothly. One can speculate about the applicability of a two-layer network for the description of functional interactions between the hemispheres. As pointed in \cite{hemi}, the presence of open 3-motifs in each hemisphere is crucial for the effective informational flow inside the hemisphere, while the links between hemispheres are evidently important for the entire brain functioning. The existence of the phase transition in the abstract two-layer dynamic network considered in \cite{maslovpt}, allows one to suggest that in functional brain networks the competition between in-layer and cross-layer interactions occurs either as a sharp 1st order phase transition which might be associated with the brain disease, or as a smooth crossover.

A bunch of evident questions concerns the observed criticality of the connectome, many of them  have been 
already posed in the literature. The most immediate question concerns the role of the long-wave excitations
intrinsic for the critical regime in the brain functioning. To make the problem more tractable one  can apply in the full power the machinery
used for the analysis of the critical regime. For example, the  fractal dimension and spectral dimension can be evaluated. 
Most of the results concerning the criticality conjecture deal with the avalanches in the 
neuron spiking phenomena ( not complete list of the references includes \cite{spikes, criticality3, criticality4,
criticality5,criticality6,criticality7}). Our results imply that  the structural connectome organization supports the critical behavior of the neural excitations.

The abstract two-layer network developing via Maslov-Sneppen rewiring algorithm, with in-layer and cross-layer interactions, demonstrates also a kind of synchronization behavior. Increasing the energy of in-layer motifs \emph{in one layer only}, we can force the clustering in \emph{both} layers simultaneously. Such a synchronization is the consequence of joint conservation laws in the network rewiring. It is worth noting that the multi-layer networks can demonstrate a bunch of new critical phenomena \cite{abrupt1, abrupt2}, such as collective phase transitions (see \cite{arenasrev} for the review) which are absent in single-layer networks. In the forthcoming works we plan to discuss statistical properties of multi-layer networks in the connectome context.

Understanding the interplay between the spectral properties of the structural connectome and the informational capacity of the brain ia also crucial for the consciousness problem \cite{88}. The discussions on this topic deals with the concept of the "integrated information" proposed in \cite{tononi}. Though its initial formulations were of very limited practical applicability the last refinements of this theory \cite{newtononi} use the standard tools of the spectral analysis.
of random networks, which makes the evaluation of the integrated information more tractable. Besides, it seems highly desirable to interpret the aspects of consciousness using the standard notions of statistical and quantum physics, such as the entanglement entropy, entanglement negativity and complexity -- see, \cite{cardy} for the review.
It is highly likely that the free energy extremization discussed in \cite{friston} could be of use for the description of the brain organization. 

Since the clustered networks allow the natural embedding into the hyperbolic geometry \cite{kryukov}, it might be possible to use modern holographic approach for the evaluation of the entanglement entropy \cite{ryu} and the complexity \cite{susskind1, susskind2} via the geometry of the hyperbolic space. Some of these ideas have been already implemented in \cite{bianconi1, bianconi2} in terms of random networks of special architecture. We believe that the progress in the field of brain studies lies at the edge of spectral theory, statistical mechanics of complex entangled systems and holography. Some initial discussion concerning the possible interplay between the
criticality of the connectome and the holographic approach can be found in \cite{dvali}.

\begin{acknowledgments}
We are grateful to A. Kamenev for the important comments.
The work of V.A. was supported within frameworks of the state task for ICP RAS 0082-2014-0001 (state registration AAAA-A17-117040610310-6). S.N. is grateful to RFBR grant 18-23-13013 for the support. The work of A.G. was performed at the Institute for Information Transmission Problems with the financial support of the Russian Science Foundation (Grant No.14-50-00150). N.P. and O.V. acknowledge the support of the RFBR grant 18-29-03167. 
O.V. thanks Basis Foundation Fellowship for the support. A.G. thanks SCGP at Stony Brook University and KITP at University of California, Santa Barbara, for the hospitality. 
\end{acknowledgments}

\begin{appendix}

\section{Spectral density of networks and motif-driven network evolution}

We have mentioned already that in constrained Erdos-Renyi networks (CERNs) with stochastic rewiring, the clustering occurs when the evolving network tends to increase the number of closed 3-motifs (triangles), $n_{\triangle}$. The concentration of triangles is fixed by the chemical potential, $\mu$. Imposing the condition of the vertex degree conservation in course of network rewiring, together with the condition of maximization the number of closed 3-motifs, one forces the network to clusterize respecting the conservation laws. The detailed analysis of phase transitions in CERN has been carried out in \cite{crit2} where it has been found that the condition of maximization of number of closed triads forces the random network with the conserved vertex degree to form a multi-clique ground state. The typical phase diagram accompanied by the visualization of the network structures is shown in \fig{fig:01} of Section \ref{s:01-1}. At $\mu=\mu_c$ the network experiences the first order phase transition and splits in the collection of weakly connected clusters.

\begin{figure}[ht]
\centerline{\includegraphics[width=16cm]{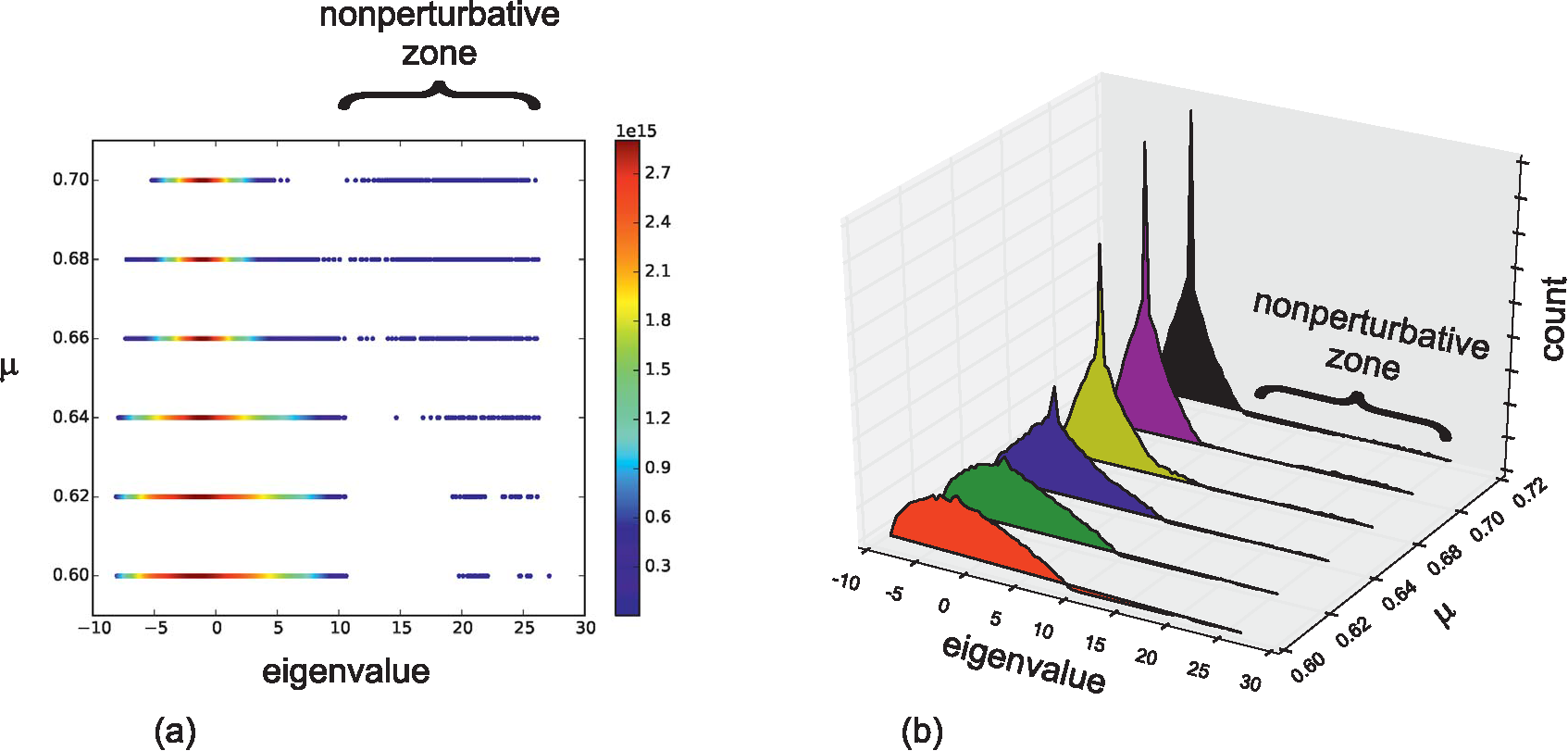}}
\caption{(a) The spectral density of ensemble of constrained Erdos-Renyi graphs for various chemical potentials $\mu$ of closed 3-motifs; (b) The same as (a) in a three-dimensional representation. The numerical results are obtained for the ensembles of 50 Erdos-Renyi graphs of 256 vertices and the bond formation probability $p=0.08$.}
\label{fig:08app}
\end{figure}

The structure of clusters (cliques) was carefully studied in \cite{crit2} via the spectral analysis of the matrix $A$ of the network. It has been shown that at $\mu<\mu_c$, the spectral density has the shape typical for Erdos-Renyi graphs with moderate connection probability, $p=O(1)<1$, being the Wigner semicircle with a single isolated eigenvalue apart. At $\mu_c$ the eigenvalues decouple from the main core and a collection of isolated eigenvalues forms the second (nonperturbative) zone as it is shown in \fig{fig:08app}a. The number of isolated eigenvalues coincides with the number of clusters formed above $\mu_c$. Averaging over ensemble of graphs patterns smears the distribution of isolated eigenvalues in the second zone. Above $\mu_c$ the support of the spectral density in the first (central) zone shrinks and the second zone becomes dense and connected. The modes in the second zone are all localized, while the ones in the central zone remain delocalized. The evolution of the spectral density of the entire network is depicted in \fig{fig:08app}b. The numerical results on spectral density evolution are obtained for the ensembles of 50 Erdos-Renyi graphs of 256 vertices each and the bond formation probability $p=0.08$.

Two important properties of the spectral density of adjacency matrices of constrained Erdos-Renyi networks have to be mentioned:
\begin{itemize}
\item The spectral densities of each cluster (clique) and of the whole network are very different \cite{crit2}. The spectrum of a clique is discrete, while the spectrum of the whole network has a two-zonal structure with the continuous triangle-shape form of the first (central) zone. We have interpreted this effect as the collectivization (or synchronization) between the modes in different clusters.
\item It was found in \cite{crit3} that there is a memory of the spectrum in the central zone on the initial state (on the preparation conditions), which is the signature of the non-ergodic nature of the system and of the presence of some hidden conservation laws.
\end{itemize}

\end{appendix}

\end{document}